\documentclass[reprint, aps, pra, floatfix]{revtex4-2}
\usepackage[dvips]{graphicx}
\usepackage[english]{babel}
\usepackage{amsmath}
\usepackage{amssymb}
\usepackage{bm}
\usepackage{color}
\usepackage{subfigure}
\usepackage{soul, xcolor}
\setstcolor{red}
\usepackage[colorlinks, citecolor = black, linkcolor = black, urlcolor = blue]{hyperref}

\newcommand{\nn}{\nonumber}
\newcommand{\p}{\partial}

\newcommand{\td}{\tilde{\delta}}
\def\ket#1{|#1\rangle}
\def\bra#1{\langle#1|}
\def\inner#1#2{\langle#1|#2\rangle}

\begin{document}
\title{Geometric phase and multipartite entanglement of Rydberg atom chains}
\author{Chang-Yan Wang}
\email{changyanwang@tsinghua.edu.cn}
\affiliation{Institute for Advanced Study, Tsinghua University, Beijing 100084, China}

\begin{abstract}
  We investigate the behavior of geometric phase (GP) and geometric entanglement (GE), a multipartite entanglement measure, across quantum phase transitions in Rydberg atom chains. Using density matrix renormalization group calculations and finite-size scaling analysis, we characterize the critical properties of transitions between disordered and ordered phases. Both quantities exhibit characteristic scaling near transition points, with the disorder to $Z_2$ ordered phase transition showing behavior consistent with the Ising universality class, while the disorder to $Z_3$ phase transition displays distinct critical properties. We demonstrate that GP and GE serve as sensitive probes of quantum criticality, providing consistent critical parameters and scaling behavior. A unifying description of these geometric quantities from a quantum geometry perspective is explored, and an interferometric setup for their potential measurement is discussed. Our results provide insights into the interplay between geometric phase and multipartite entanglement near quantum phase transitions in Rydberg atom systems, revealing how these quantities reflect the underlying critical behavior in these complex quantum many-body systems.
\end{abstract}
\maketitle

Rydberg atom arrays have emerged as a powerful platform for studying quantum many-body physics \cite{bernien_probing_2017, browaeys_manybody_2020, ebadi_quantum_2021}, offering a high degree of controllability and the ability to study a wide range of physical phenomena, such as quantum many-body scars \cite{turner_weak_2018, turner_quantum_2018, lin_exact_2019, lin_quantum_2020, serbyn_quantum_2021, turner_correspondence_2021, cheng_tunable_2022, pan_composite_2022, yao_quantum_2022, chandran_quantum_2023}, quantum spin liquid \cite{whitsitt_quantum_2018, verresen_prediction_2021, semeghini_probing_2021, giudice_trimer_2022, giudici_dynamical_2022, lee_landauforbidden_2023, samajdar_emergent_2023, slagle_quantum_2022, verresen_unifying_2022,  yan_triangular_2022, cheng_variational_2023, kalinowski_nonabelian_2023, kornjaca_trimer_2023, tarabunga_gaugetheoretic_2022} and Kibble-Zurek mechanism \cite{keesling_quantum_2019, chepiga_kibblezurek_2021}. These systems also have attracted significant attention due to their potential applications in quantum information processing and quantum computation \cite{adams_rydberg_2019, cohen_quantum_2021, cong_hardwareefficient_2022, evered_highfidelity_2023, nguyen_quantum_2018, paredes-barato_alloptical_2014, ryabtsev_applicability_2005, saffman_quantum_2010, saffman_quantum_2016, wu_concise_2021, zhao_rydbergatombased_2017, bluvstein_controlling_2021, ebadi_quantum_2022, omran_generation_2019}. Understanding the quantum phases and phase transitions in Rydberg atom arrays is crucial for harnessing their capabilities and exploring fundamental questions in quantum physics.

The geometric phase and quantum entanglement have proven to be powerful tools for characterizing quantum critical behavior in many-body systems over the past decades \cite{osterloh_scaling_2002, carollo_geometry_2020, chiara_genuine_2018}. The geometric phase, originally introduced by Berry \cite{berry_quantal_1984}, has proven invaluable in studying topological properties of quantum systems, such as the quantum Hall effect \cite{thouless_quantized_1982}. It has also been instrumental in investigating quantum critical phenomena in various systems, including the $XY$ model \cite{zhu_scaling_2006, carollo_geometric_2005, hamma_berry_2006, pachos_geometric_2006, patra_pathdependent_2011, quan_finitetemperature_2009, reuter_geometric_2007} and the Dicke model \cite{chen_critical_2006, plastina_scaling_2006}, etc. Concurrently, geometric entanglement \cite{wei_geometric_2003, barnum_monotones_2001, shimony_degree_1995}, a measure of multipartite entanglement \cite{chiara_genuine_2018}, has been successfully applied to study critical phenomena and detect phase transitions \cite{wei_global_2005, orus_universal_2008, montakhab_multipartite_2010, hofmann_scaling_2014, orus_visualizing_2010, radgohar_global_2018}. Mousolou et al. \cite{azimimousolou_unifying_2013} proposed a unifying description that combines these two quantities into a complex-valued geometric entanglement, using the $XY$ model as an illustrative example. 

In this work, we investigate the behavior of GP and GE in Rydberg atom chains across quantum phase transitions. We employ density matrix renormalization group (DMRG) method \cite{perez-garcia_matrix_2007, schollwock_densitymatrix_2005, schollwock_densitymatrix_2011, white_density_1992} to compute these quantities for various system sizes. By applying finite-size scaling analysis, we extract quantum phase transition points and critical exponents for the quantum phase transitions from disorder to $Z_2$ ordered phase and from disorder to $Z_3$ ordered phase, demonstrating the effectiveness of these geometric quantities as probes of quantum criticality in Rydberg atom systems.

We also discuss the unifying description of GP and GE based on a quantum geometry perspective adopted from previous work in the $XY$ model \cite{azimimousolou_unifying_2013}. This approach provides a deeper understanding of the connection between these quantities in the context of quantum phase transitions. Additionally, we explore the applicability of an interferometric setup, also proposed in the context of the $XY$ model \cite{azimimousolou_unifying_2013}, for the potential experimental measurement of these geometric quantities in Rydberg atom chains.

The rest of this paper is organized as follows: in Sec. \ref{sec:rydberg}, we review the Hamiltonian of the Rydberg atom chains, and their phases; In Sec. \ref{sec:gp}, we review the geometric phase and why it can be used to characterize quantum phase transitions; in Sec. \ref{sec:ge}, we discuss two different methods to calculate the geometric entanglement; in Sec. \ref{sec:numerics}, we present our numerical result for GP and GE and extract the critical points and critical exponent using finite size scaling; in Sec. \ref{sec:unify} and Sec. \ref{sec:interf}, we discuss the relation of GP and GE based on a quantum geometry perspective and an interferometer setup; we conclude in Sec. \ref{sec:conclusion}.

\section{Hamiltonian of Rydberg atom chains}\label{sec:rydberg}
We consider the Hamiltonian of 1d Rydberg atom chain with length $L$ in open boundary condition \cite{bernien_probing_2017, zeiher_coherent_2017}
\begin{eqnarray}\label{ham}
  H = \sum_{i} (\Omega \hat{\sigma}_i^x - \delta \hat{n}_i) + \sum_{i < j} V(|i-j|) \hat{n}_i \hat{n}_j,
\end{eqnarray}
where $\hat{\sigma}_i = |r_i\rangle\langle g_i| + |g_i\rangle\langle r_i|$ describes the couple between the internal atomic ground state $|g_i\rangle$ and a Rydberg excited state $|r_i\rangle$ of the $i$th atom and $\hat{n}_i = |r_i\rangle\langle r_i|$. $\Omega$ and $\delta$ denote the Rabi frequency and detuning of the coherent laser driving, respectively. The term $V(|i-j|) = C_6/|i-j|^6$ describes the van der Waals interaction between the $i$th and $j$th atoms when both are excited to the Rydberg state.

\begin{figure}
  \includegraphics[width = .3\textwidth]{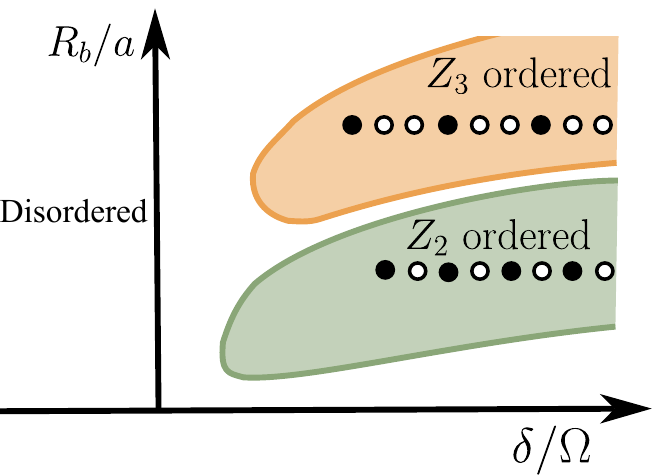}
  \caption{Schematics of the phase diagram of the 1d Rydberg atom chain. As the blockade radius increases, the $Z_2$ order phase emerges as the occupation of the Rydberg state in next-neighbor atoms is suppressed, then $Z_3$ order phase emerges as occupation of the Rydberg state in next-next-neighbor atoms is further suppressed.}
  \label{fig:phase_diag}
\end{figure}

The long-range van der Waals interaction can be parametrized by the Rydberg blockade radius $R_b$, defined as $V(R_b/a) \equiv \Omega$, where $a$ is the lattice spacing. When $R_b/a$ becomes sufficiently large, the long-range van der Waals interaction disfavors the nearest-neighbor atoms being excited to the Rydberg state, leading to the formation of the $Z_2$ ordered phase. As $R_b/a$ further increases, the occupation of the Rydberg state in next-nearest-neighbor atoms is also suppressed, resulting in the formation of the $Z_3$ ordered phase \cite{bernien_probing_2017}, as shown in Fig. \ref{fig:phase_diag}.

\section{Geometric phase of the Rydberg array}\label{sec:gp}
The geometric phase, or Berry phase, is a phase factor acquired by a quantum system undergoing an adiabatic cyclic evolution in parameter space. It arises due to the non-trivial topology of the system's Hilbert space and the curvature associated with the parameter space. Geometric phase can profoundly affect physical observables, such as energy levels, interference patterns, and quantum transport properties, leading to phenomena like the Aharonov-Bohm effect and the quantum Hall effect \cite{thouless_quantized_1982}. These effects have initiated extensive studies of topological phases of matter \cite{hasan_colloquium_2010, qi_topological_2011}. Moreover, the geometric phase can be used to characterize quantum phase transitions, and critical exponents can be extracted from the scaling behavior of geometric phases.

When a quantum system's parameters, denoted by $\mathbf{R}$, are varied slowly and cyclically, its eigenstate $|\psi_n(\mathbf{R}(t))\rangle$, with $t \in [0, 1]$ parameterizing the adiabatic path, acquires a phase factor upon returning to its initial state: $|\psi_n(\mathbf{R}(1))\rangle = e^{i\Phi_B}|\psi_n(\mathbf{R}(0))\rangle$, where $\Phi_B$ is the geometric phase. The geometric phase can be calculated by integrating the Berry connection $A_B$ along the adiabatic path $C$ in parameter space:
\begin{eqnarray} \label{berry}
\Phi_B = \int_0^1 A_B dt, \ A_B = i\langle\psi_n(\mathbf{R}(t))|\frac{d}{dt}|\psi_n(\mathbf{R}(t))\rangle.
\end{eqnarray}
We define the GP density as $\phi_B = \frac{\Phi_B}{L}$ for later use.

To calculate the geometric phase, one must construct an adiabatic path. For example, in the $XY$ model, a natural choice is spin rotation along the $z$-axis. Similarly, we can construct such an adiabatic path for the Rydberg atom chain by treating the ground state $|g_i\rangle$ and the Rydberg excited state $|r_i\rangle$ of each atom as a pseudo-spin. Defining $\hat{\sigma}_i^z = 2|r_i\rangle\langle r_i| - 1$, we have the adiabatic Hamiltonian as:
\begin{eqnarray}
H(t) = U(t)H U^\dag(t), \ U(t) = \prod_{i=1}^L e^{-i t \pi \hat{\sigma}_i^z }.
\end{eqnarray}
With this adiabatic path, the adiabatic ground state is given by:
\begin{eqnarray}
|\psi(t)\rangle = U(t)|\psi_0\rangle,
\end{eqnarray}
where $|\psi_0\rangle$ is the ground state of $H$ when the adiabatic parameter $t = 0$. One can then calculate the geometric phase according to Eq.(\ref{berry}). Unlike the $XY$ model, which is exactly solvable using the Jordan-Wigner transformation, the Rydberg atom array is not exactly solvable even in 1D in general \cite{park_graphtheoretical_2025}. The Rydberg atom chain actually can be regarded as an Ising model with long-range interactions in the presence of both transverse and longitudinal fields \cite{lesanovsky_manybody_2011}. The presence of both transverse and longitudinal fields, along with long-range interactions, prevents the analytical solution of the Rydberg atom chain. Thus, one needs to use sophisticated numerical methods such as the DMRG method.

Before proceeding to the numerical results, we shall briefly explain why the geometric phase can characterize quantum critical behavior. Quantum phase transitions are associated with level-crossing, and the singularity of the geometric phase indeed results from such level-crossing. The geometric phase along a closed path $C$ in parameter space can also be given by the integration of the Berry curvature over the region $S$ enclosed by this path: $\Phi_B = -\int_S \mathbf{F}(\mathbf{R}) \cdot d\mathbf{S}$, where $d\mathbf{S}$ denotes the area element of region $S$, and the Berry curvature for the ground state $|\psi_0\rangle$ is given by \cite{berry_quantal_1984}:
\begin{eqnarray}
\mathbf{F}(\mathbf{R}) = \mathrm{Im}\sum_{n\neq 0}\frac{\langle \psi_0|\nabla_\mathbf{R} H|\psi_n\rangle \times \langle \psi_n|\nabla_\mathbf{R} H|\psi_0\rangle}{(E_n - E_0)^2},
\end{eqnarray}
where $|\psi_n(\mathbf{R})\rangle$ is the $n$th eigenstate of the Hamiltonian, i.e., $H(\mathbf{R})|\psi_n(\mathbf{R})\rangle = E_n(\mathbf{R})|\psi_n(\mathbf{R})\rangle$. Hence, we can see that when there is level-crossing around the ground state energy, the Berry curvature diverges. As the quantum phase transition occurs at level crossings or avoided level crossings, thus, GP can capture such quantum phase transition.

\section{Multipartite Entanglement of the Rydberg array}\label{sec:ge}
There are several multipartite entanglement measures; the geometric entanglement (GE) is one of them. The idea of GE is based on the observation that the multipartite entanglement of a quantum 
state $|\psi\rangle$ can be captured by the overlap between this state and its closest product state, i.e. 
\begin{eqnarray}\label{lambda_max}
  \Lambda_\mathrm{max} = \mathrm{max}_{\psi_p}|\langle \psi_p|\psi\rangle|.
\end{eqnarray}
For the $XY$ model, the quantity $\Lambda_\mathrm{max}$ is actually the entanglement eigenvalue of the $XY$ ground state. The GE of the state $|\psi\rangle$ is defined as 
\begin{eqnarray}
  E(\psi) = -\log_2 \Lambda_\mathrm{max}^2.
\end{eqnarray}
And the GE density is defined as $\mathcal{E}(\psi) = E(\psi)/L$. 

\begin{figure}
  \includegraphics[width = 0.48\textwidth]{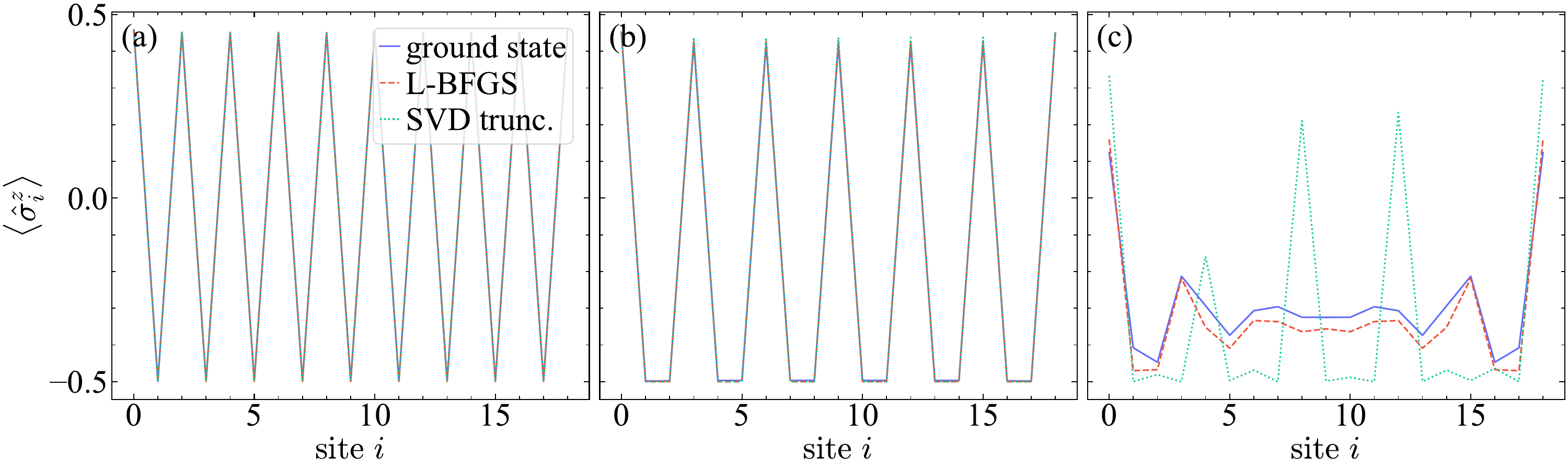}
  \caption{$\langle\hat{\sigma}_i^z \rangle$ for the exact ground state calculated from the DMRG method and the product states calculated through the L-BFGS and SVD truncation methods respectively for a Rydberg chain of length $L = 19$ in different phases. (a) $Z_2$ ordered state with $\delta/\Omega = 2.5, R_b/a = 1.5$, (b) $Z_3$ ordered state with $\delta/\Omega = 2.5, R_b/a = 2.5$. (c) Disordered state with $\delta/\Omega = 1.5, R_b/a = 2.5$.}
  \label{fig:ent_compare}
\end{figure}

It has been demonstrated that the GE of the ground state in the $XY$ model exhibits features characteristic of quantum criticality. These features include universality, critical exponents, and scaling behavior, which are manifested through the singular behavior of the GE density. By performing scaling analysis on the divergences observed at the singular points of the GE density, critical exponents corresponding to different universality classes have been extracted.

Due to the realness of the Hamiltonian of the Rydberg atom chains, we can assume the closest product state to the ground state $|\psi\rangle$ of the Hamiltonian takes the following form
\begin{eqnarray}
  |\psi_p(\bm{\theta})\rangle = \otimes_{i=1}^L \Big[\cos\left(\theta_i\right) |r_i\rangle \pm \sin\left(\theta_i\right) |g_i\rangle\Big],
\end{eqnarray}
where $\bm{\theta} = \{\theta_1, \theta_2, ..., \theta_L\}$ are the parameters to be determined to maximize the overlap i.e. $\Lambda_\mathrm{max} = \mathrm{max}_{\bm{\theta}} |\langle \psi_p(\bm{\theta})|\psi\rangle|$. In previous studies, such as for the $XY$ model, translational symmetry is always adopted, hence, one can take $\theta_i = \theta$, which can simplify the calculation. However, such an ansatz for the closest product state is not valid for the $Z_2$ and $Z_3$ ordered states of the Rydberg atom chain, as these states are no longer translationally invariant. Therefore, we cannot simply set $\theta_i = \theta$, instead, we need to numerically minimize the function $\mathcal{F}(\bm{\theta}) = -|\inner{\psi_p(\bm{\theta})}{\psi}|$ over possible $\bm{\theta}$, which includes as many as hundreds of parameters in our case. Such a task can be achieved by the so-called limited memory Broyden-Fletcher-Goldfarb-Shanno (L-BFGS) method \cite{nocedal_numerical_2006}, which is a large scale optimization method that can minimize a function with many parameters. This algorithm has been implemented in software packages such as \texttt{Optim.jl} \cite{mogensen_optim_2018}.

An alternative approach to calculating the closest product state involves a two-step process based on the DMRG method \cite{schneider_entanglement_2022}. First, the exact matrix product state (MPS) ground state is computed using DMRG calculations with a high bond dimension. Then, a singular value decomposition (SVD) is performed on the MPS, and it is truncated to a single dominant singular value, reducing the MPS to a product state. More specifically, for the MPS ground state, the SVD is given by
\begin{eqnarray}
  \ket{\Psi_\mathrm{MPS}} = \sum_i \lambda_i \ket{\psi_i^L}\bra{\psi_i^R}.
\end{eqnarray}
Then we keep only the contribution from the largest singular value $\lambda_1$, i.e. $\ket{\psi_p} = \lambda_1 \ket{\psi_1^L}\bra{\psi_1^R}$. We can repeat this process for each bond of the MPS state to obtain the product state $\ket{\psi}$ that is closest to the MPS ground state $\ket{\Psi_\mathrm{MPS}}$. Subsequently, this obtained product state undergoes several variational optimization sweeps, where the bond dimension is fixed at unity, until the overlap with the exact ground state converges.

To compare these two algorithms, in Fig. \ref{fig:ent_compare}, we show the expected value of the operator $\hat{\sigma}_i^z$ for the exact ground state calculated from the DMRG method and the product states calculated through the L-BFGS and SVD truncation methods respectively for a Rydberg chain of length $L = 19$ in different phases. In Fig. \ref{fig:ent_compare} (a) and (b), we show $\langle\hat{\sigma}_i^z \rangle$ for the $Z_2$ and $Z_3$ ordered states respectively. In these two phases, $\langle\hat{\sigma}_i^z \rangle$ of the product states obtained from the L-BFGS method and the SVD truncation method almost coincide with that of the exact ground state. However, for the disordered phase as shown in Fig. \ref{fig:ent_compare} (c), $\langle\hat{\sigma}_i^z \rangle$ of the product state calculated from L-BFGS method is closer to that of the exact ground state that the SVD truncation method. Thus, we will adopt the L-BFGS method in our numerical calculations.

\section{Numerical results}\label{sec:numerics}

Due to the lack of an analytical solution for the Hamiltonian Eq.(\ref{ham}), we employ the state-of-the-art numerical method, specifically the density matrix renormalization group method, to determine the ground state using the software \texttt{ITensors.jl}. The source code and data are available online \cite{wang_codegeometric_2025}. Although the Hamiltonian Eq.(\ref{ham}) contains long-range van der Waals interaction, we adopt a truncation scheme for the parameter range $R_b/a \lesssim 3$ considered here. We set the van der Waals interaction to zero for atoms separated by more than $5a$, i.e., $V(|i-j|>5) = 0$. In our numerical calculations, we set both the truncation error and relative energy tolerance error to $10^{-11}$.

\begin{figure}
  \includegraphics[width = .48\textwidth]{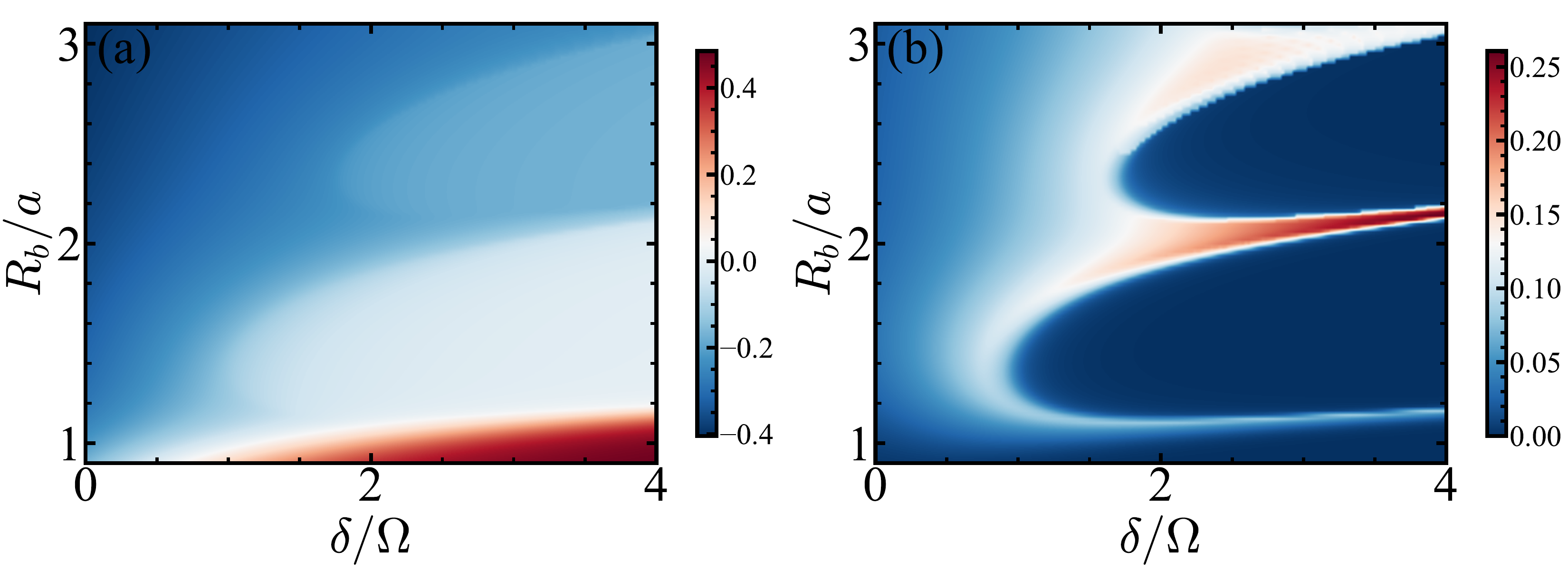}
  \caption{(a) GP density of the 1d Rydberg atom chain. (b) GE density of the 1d Rydberg atom chain. The system size is $L=121$.}
  \label{fig:berry_ent}
\end{figure}

Fig. \ref{fig:berry_ent} illustrates the geometric phase density $\phi_B$ and the geometric entanglement density $\mathcal{E}$ as functions of the detuning $\td = \delta/\Omega$ and the Rydberg blockade radius $R_b/a$, for a Rydberg chain of length $L=121$. Both quantities clearly exhibit lobe shapes, corresponding to the $Z_2$ and $Z_3$ ordered states. This demonstrates that the geometric phase and geometric entanglement can characterize the phase of the Rydberg chain without prior knowledge of the system's symmetry. Furthermore, as we will demonstrate, analyzing the scaling behavior of the derivatives of GP and GE density with respect to $\td$ allows us to extract the critical exponents of the phase transition in the Rydberg chain.

We define the derivative of the GP density with respect to the tuning parameter $\td = \delta/\Omega$ as:
\begin{eqnarray}
\phi_B' = \frac{\partial \phi_B}{\partial \td}
\end{eqnarray}
and the derivative of the GE density as:
\begin{eqnarray}
\mathcal{E}' = \frac{\partial \mathcal{E}}{\partial \td}
\end{eqnarray}

\begin{figure}
  \includegraphics[width = .45\textwidth]{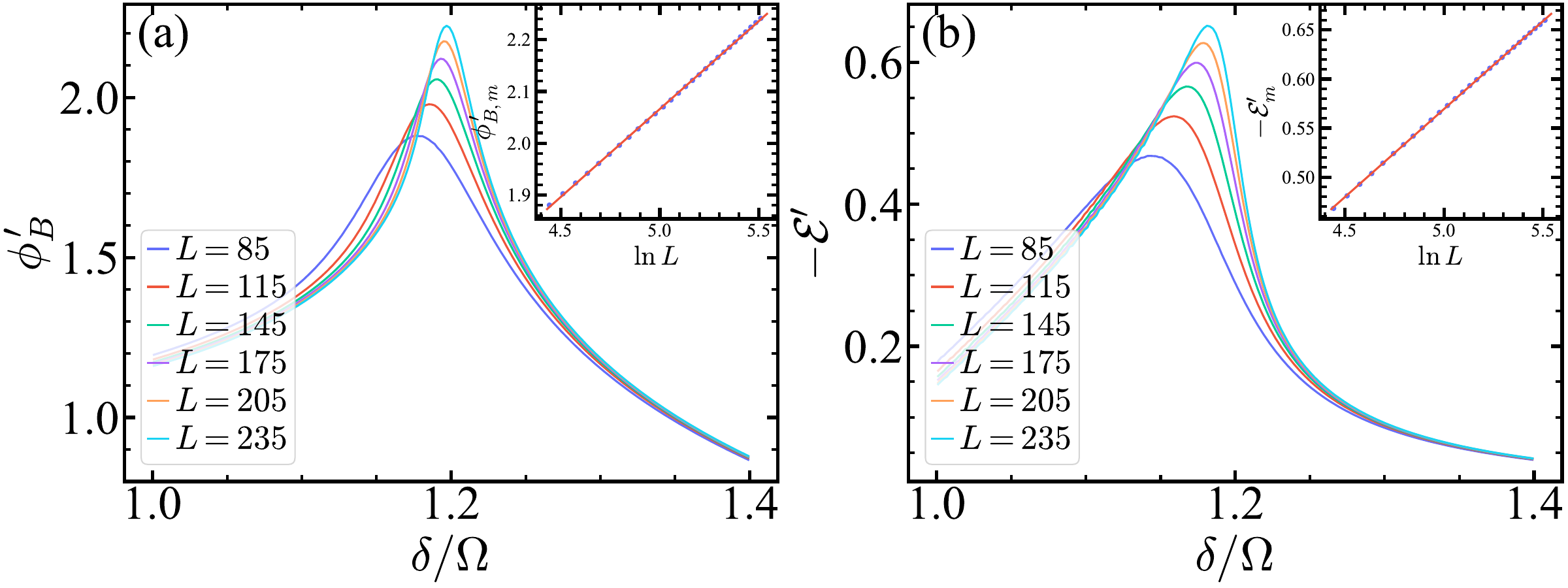}
  \caption{(a) Derivative of the GP density, denoted as $\phi_B'$ over the parameter $\delta/\Omega$, when $R_b/a = 1.6$. Inset: max value of $\phi_B'$ for different system size L. (b) Derivative of the GE density, denoted as $\mathcal{E}'$ over the parameter $\delta/\Omega$, when $R_b/a = 1.6$. Inset: max value of $\mathcal{E}'$ for different system size L. }
  \label{fig:ber_geo_1.6}
\end{figure}

Fig. \ref{fig:ber_geo_1.6} presents $\phi_B'$ and $-\mathcal{E}'$ as functions of the tuning parameter $\td$ for different system sizes $L$, at $R_b/a = 1.6$. For finite system sizes, we observe no true divergence, but rather a peak for each system size. The maximum of each peak defines a pseudo-critical point $\delta_m$, which approaches the real critical point $\delta_c$ as the system size increases. The insets show the values of $\phi_B'$ and $-\mathcal{E}'$ at the pseudo-critical points as functions of the logarithm of system size $L$. Both $\phi_{B,m}'$ and $-\mathcal{E}_m'$ increase linearly with $\ln L$. Specifically, for $\phi_{B,m}'$, we have:
\begin{eqnarray}\label{phi_bm_log}
\phi_{B,m}' = \kappa_{1,\phi'} \ln L + \mathrm{const.}
\end{eqnarray}
By using least-square fitting, we have $\kappa_{1,\phi'} = 0.340$ for the GP density derivative, and $\kappa_{1,\mathcal{E}'} = 0.179$ for the GE density derivative when the tuning parameter $\td = 1.6$. This logarithmic behavior at the pseudo-critical points suggests a logarithmic behavior for the GP derivative around the critical point $\td_c$ in the thermodynamic limit \cite{barber_phase_1983}:
\begin{eqnarray}\label{phi_b_log}
\phi_B' = \kappa_{2,\phi'} \ln|\td - \td_c| + \mathrm{const.}
\end{eqnarray}
The critical exponent $\nu$ can be obtained from $\nu = |\kappa_{2,\phi'} / \kappa_{1,\phi'}|$ \cite{barber_phase_1983}(See Appendix \ref{fss} for more detail). For systems like the $XY$ model where $\kappa_2$ can be calculated analytically, this method can directly extract the critical exponent.

However, for the Rydberg chain, $\kappa_2$ cannot be calculated analytically. Therefore, we must employ the finite-size scaling method to extract the critical exponent. For quantities with logarithmic scaling behavior like $\phi_B'$ (as in Eq. (\ref{phi_b_log})), it can be shown that the data for $F_{\phi_B'} = 1-\exp(\phi_B' - \phi_{B,m}')$ as a function of $L^{1/\nu}(\td - \td_c)$ should collapse onto a single curve, independent of system size $L$ \cite{barber_phase_1983}. The challenge lies in not knowing the function of this curve a priori. To quantify the quality of data collapse, we adopt the quantity $S(\td_c, \nu)$ defined in Ref. \cite{houdayer_lowtemperature_2004} (See Appendix \ref{app:quality} for more detail). This quantity measures the mean square distance between the data and the fitted curve for various values of $\delta_c$ and $\nu$. By minimizing $S(\td_c,\nu)$ over possible values of $\delta_c$ and $\nu$, we can determine the critical point and critical exponent. Since the GE density $\mathcal{E}'$ also exhibits a logarithmic scaling behavior, we can define the quantity $F_{\mathcal{E}'} = 1 - \exp(\mathcal{E}' - \mathcal{E}_m')$. We can use the same finite size scaling method to find the best data collapse of $F_{\mathcal{E}'}$ as a function of $L^{1/\nu}(\td - \td_c)$ to extract the quantum critical point and critical exponent.

\subsection{Disorder to $Z_2$ ordered phase transition}

% In Fig. \ref{fig:fss_1.6}, we show the data collapse for the quantity $F_{\phi'}$ and $F_{\mathcal{E}'}$ as a function of $L^{1/\nu}(\td - \td_c)$, when $R_b/a = 1.6$. In this case, a phase transition form disorder to $Z_2$ ordered phase can happen. We can see that the data collapse well for both cases irrespective of the system sizes. Form the data collapse of the GP density, we obtain the critical point $\td_{c,\phi'}$ = 1.020, and the critical exponent $\nu_{\phi'} = 1.059$. And for the GE density case, we obtain the critical point $\td_{c,\mathcal{E}'} = 1.018$, and the critical exponent $\nu_{c, \mathcal{E}'} = 1.067$. We can see that for both the GP and GE densities, the critical exponent is close to 1, which is compatible with that the disorder to $Z_2$ ordered phase transition belongs to the Ising universality class.

\begin{figure}
  \includegraphics[width = .45\textwidth]{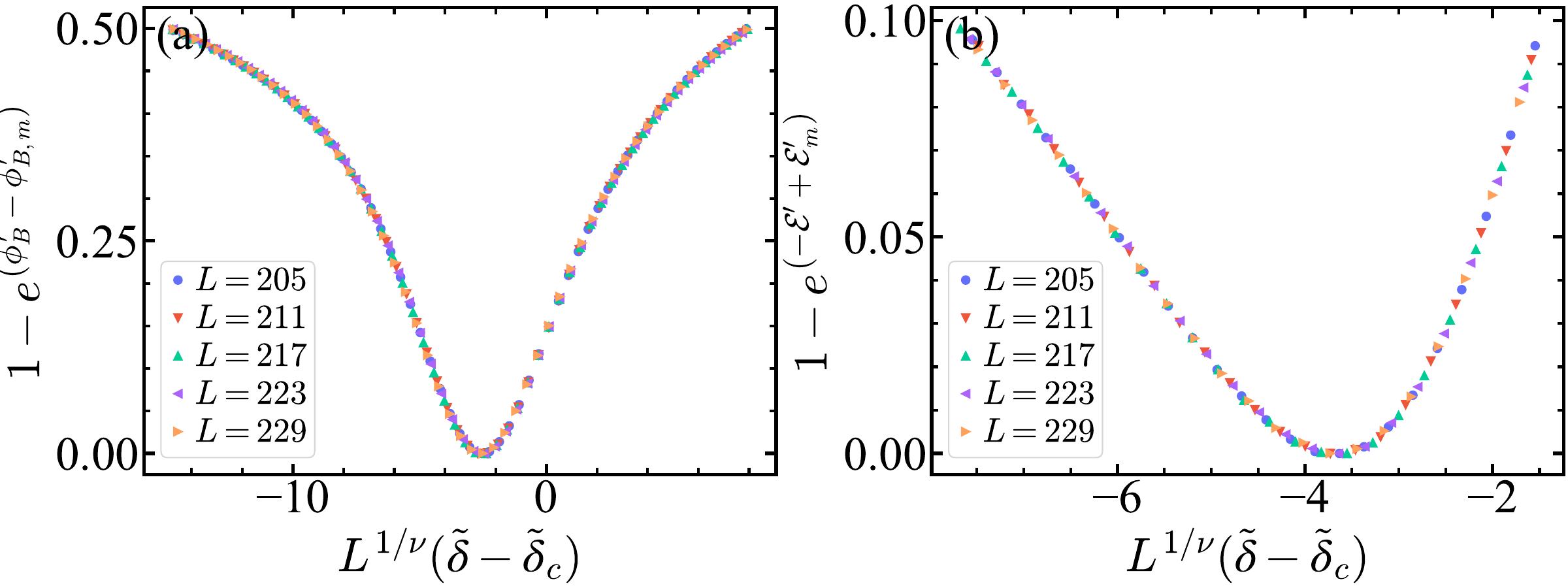}
  \caption{Data collapse for the derivative of (a) GP and (b) GE, when $R_b/a = 1.6$.}
  \label{fig:fss_1.6}
\end{figure}

Fig. \ref{fig:fss_1.6} illustrates the data collapse for quantities $F_{\phi'}$ and $F_{\mathcal{E}'}$ as functions of $L^{1/\nu}(\td - \td_c)$, for $R_b/a = 1.6$. In this regime, a phase transition from disorder to $Z_2$ ordered phase is expected. The data collapse is robust across different system sizes for both GP and GE derivatives. Analysis of the GP density data collapse yields a critical point $\td_{c,\phi'} = 1.209$ and a critical exponent $\nu_{\phi'} = 1.010$. Similarly, for the GE density, we obtain $\td_{c,\mathcal{E}'} = 1.207$ and $\nu_{c, \mathcal{E}'} = 1.093$. Notably, both GP and GE densities produce critical exponents close to 1, consistent with the expectation that the disorder to $Z_2$ ordered phase transition belongs to the Ising universality class. These results are also compatible with that in Ref. \cite{yu_fidelity_2022}. These results demonstrate the effectiveness of both GP and GE in characterizing this quantum phase transition.

\subsection{Disorder to $Z_3$ ordered phase transition}

% In Fig. \ref{fig:berry_geo_2.3}, we show the numerical results for the GP and GE density derivatives varying with the tuning parameter $\td$ for different system sizes, as well as their values in the pseudo-critical points varying with the logarithm of system size $L$, when $R_b/a = 2.3$. Here, a phase transition from disorder to $Z_3$ ordered phase can happen. From the inset of Fig. \ref{fig:berry_geo_2.3}, we can see that the values of both $\phi_{b}'$ and $-\mathcal{E}'$ at the pseudo-critical points increase linearly with the logarithm of system size $L$. Using the least-square fitting, we obtain $\kappa_{1,\phi'} = 0.106$ for $\phi_{B}'$ and $\kappa_{1,\mathcal{E}'} = 0.027$ for $-\mathcal{E}'$. 

\begin{figure}
  \includegraphics[width = .45\textwidth]{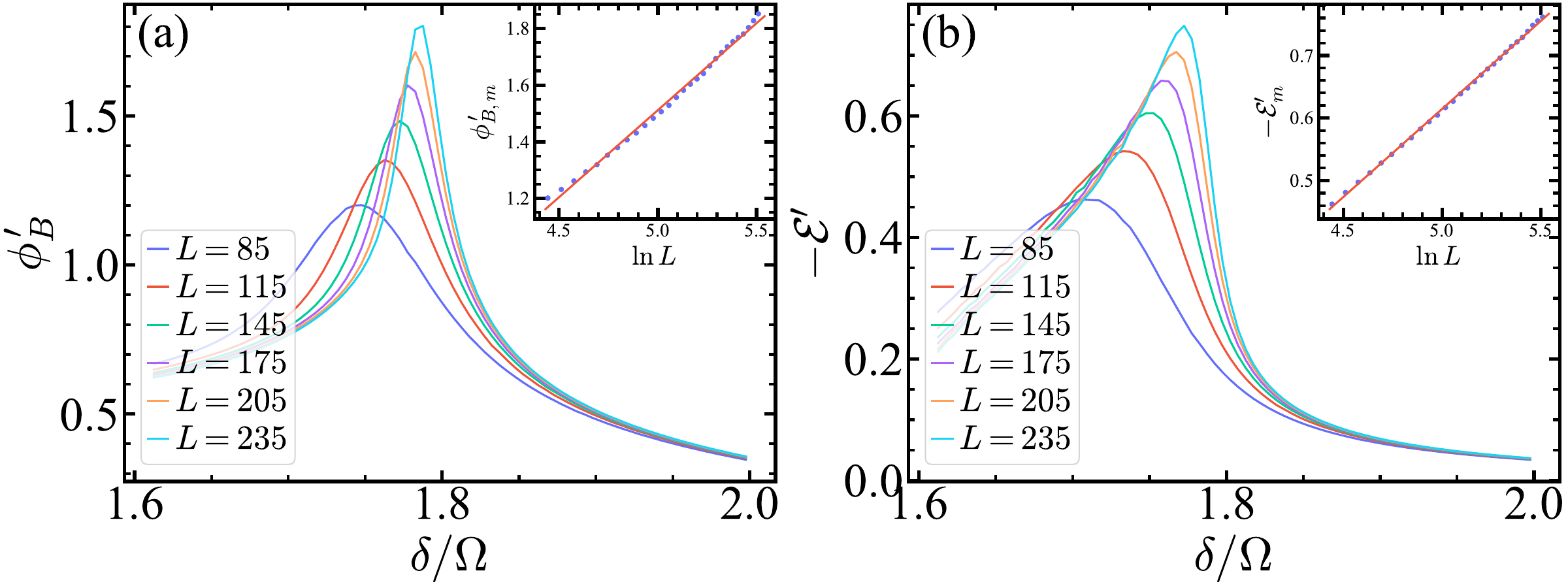}
  \caption{Derivative of (a) the GP density $\phi_B'$ and (b) the GE density over the tuning parameter $\td = \delta/\Omega$, when $R_b/a = 2.3$. Inset: The values at pseudo-critical points of (a) $\phi_B'$ and (b) $\mathcal{E}'$ varying with the logarithm of system size $L$.}
  \label{fig:berry_geo_2.3}
\end{figure}

% Given this logarithm scaling behavior, we can also obtain the critical point and the critical exponent using the finite size scaling method. In Fig. \ref{fig:fss_2.3}, we show the data collapse for qualities $F_{\phi'}$ and $F_{\mathcal{E}'}$ as functions of $L^{1/\nu}(\td - \td_c)$ when $R_b/a = 2.3$. From this finite size scaling analysis, we obtain the critical point $\td_{c,\phi'} = 1.797$ and the critical exponent $\nu_{\phi'} = 0.654$ using the GP density derivative data, and $\td_{c,\mathcal{E}'} = 1.799$, $\nu_{\mathcal{E}'} = 0.828$ using the GE density derivative data. We can see that the critical points $\td_{c,\phi'}$ and $\td_{c,\mathcal{E}'}$ are close to each other, they are also compatible with the critical point obtained in Ref. \cite{yu_fidelity_2022} using different approach. 

Figure \ref{fig:berry_geo_2.3} illustrates the numerical results for the GP and GE density derivatives as functions of the tuning parameter $\td$ for various system sizes, when $R_b/a = 2.3$. This regime corresponds to a potential phase transition from a disordered to a $Z_3$ ordered phase. The insets show the values of $\phi_{B}'$ and $-\mathcal{E}'$ at the pseudo-critical points plotted against the logarithm of system size $L$. Both quantities exhibit a clear linear increase with $\ln L$, consistent with the expected logarithmic scaling behavior. Using least-squares fitting, we obtain scaling coefficients $\kappa_{1,\phi'} = 0.614$ for $\phi_{B}'$ and $\kappa_{1,\mathcal{E}'} = 0.281$ for $-\mathcal{E}'$. This logarithmic scaling provides a foundation for further finite-size scaling analysis.

\begin{figure}
  \includegraphics[width = .45\textwidth]{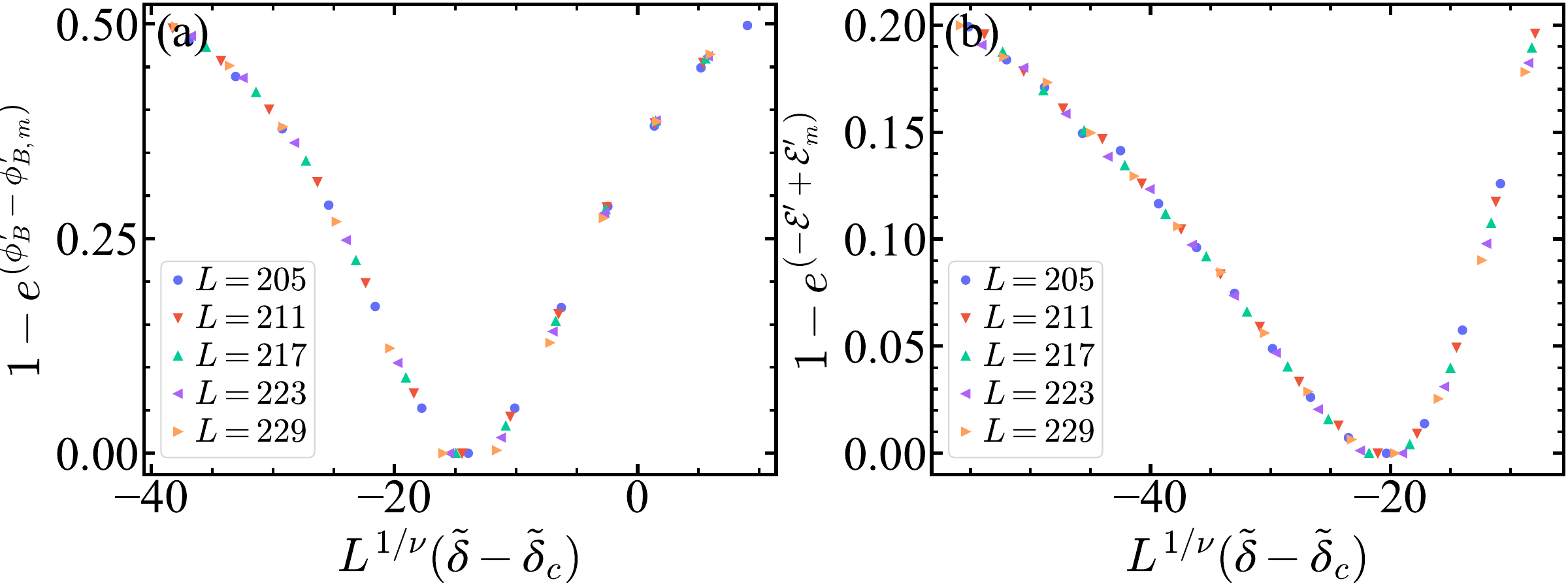}
  \caption{Data collapse for the derivative of (a) GP and (b) GE, when $R_b/a = 2.3$.}
  \label{fig:fss_2.3}
\end{figure}

Figure \ref{fig:fss_2.3} presents the data collapse for quantities $F_{\phi'}$ and $F_{\mathcal{E}'}$ as functions of $L^{1/\nu}(\td - \td_c)$ for $R_b/a = 2.3$. Employing finite-size scaling analysis on the GP density derivative data yields a critical point $\td_{c,\phi'} = 1.801$ and a critical exponent $\nu_{\phi'} = 0.802$. Similarly, analysis of the GE density derivative data produces $\td_{c,\mathcal{E}'} = 1.800$ and $\nu_{\mathcal{E}'} = 0.825$. The close agreement between $\td_{c,\phi'}$ and $\td_{c,\mathcal{E}'}$ as well as the critical exponents lends credibility to our results. Moreover, these critical points and critical exponents are consistent with those obtained in Ref. \cite{yu_fidelity_2022, samajdar_numerical_2018} using alternative methods, further validating our approach. 

\section{Unifying description of GP and GE through quantum geometry}\label{sec:unify}

At first glance, the GE and GP seem unrelated based on respective definitions. However, they can be related to each other through the so-called quantum geometric tensor \cite{azimimousolou_unifying_2013}. In this section, we shall discuss the relationship between them in the Rydberg atom chain case. 

We consider the Hilbert space $\mathcal{H}$ formed by the normalized quantum many-body states of the Rydberg atom chain of length $L$. Assuming this space $\mathcal{H}$ can be parametrized by the coordinates ${\xi_1, \xi_2, ..., \xi_{\dim\mathcal{H}}}$, We can define the quantum metric tensor for a quantum state $\ket{\psi}\in \mathcal{H}$ as \cite{provost_riemannian_1980}
\begin{eqnarray}
  \mathcal{T}_{\mu\nu} = \langle \p_\mu\psi|\p_\nu\psi\rangle - \langle \p_\mu\psi|\psi\rangle\langle \psi|\p_\nu\psi\rangle,
\end{eqnarray}
where $\p_\mu \ket{\psi} = \p\ket{\psi}/\p\xi_{\mu}$. This quantum geometric tensor comprises a symmetric real part and an antisymmetric part, i.e.
\begin{eqnarray}
  \mathcal{T}_{\mu\nu} = g_{\mu\nu} + i \mathcal{F}_{\mu\nu}.
\end{eqnarray} 
The real part $g_{\mu\nu}$ of the quantum tensor is known as the Fubini-Study metric in the Hilbert space, while the imaginary part $\mathcal{F}_{\mu\nu}$ corresponds to the Berry curvature. 

For a state $\ket{\psi}$ in the Hilbert space $\mathcal{H}$, the overlap of states and the Fubini-Study metric are related as follows \cite{provost_riemannian_1980}
\begin{eqnarray}
  |\inner{\psi(\xi)}{\psi(\xi+d\xi)}| = 1 - \frac{1}{2} ds^2,
\end{eqnarray}
where $ds^2 = g_{\mu\nu}d\xi_\mu d\xi_\nu$. Consequently, the overlap $\Lambda_\mathrm{max}$ between the state $|\psi\rangle$ and its closest product state $|\psi_p\rangle$, can be represented by the integration of $ds$ over the geodesics $\Gamma$ connecting $|\psi\rangle$ and $|\psi_p\rangle$,
\begin{eqnarray}\label{ge_tensor}
  \Lambda_\mathrm{max} = 1 - \frac{1}{2}\left[\int_\Gamma ds\right]^2.
\end{eqnarray}
To obtain the geometric phase from the imaginary part of the geometric tensor, we need to define a closed path in the space $\mathcal{H}$. Such a closed path can be described by a unitary operator, $U(s) = \otimes_i U_i(s)$, where $U_i(s)$ is a unitary operator acting on the $i$th Rydberg atom. Hence, the geometric phase can be expressed as the integration over the surface $\Sigma$ enclosed by this closed path, i.e.
\begin{eqnarray}\label{gp_tensor}
  \Phi_B = \int_\Sigma F_{\mu\nu}dS^{\mu\nu},
\end{eqnarray}
where $dS^{\mu\nu}$ is the surface element of $\Sigma$. 

To further understand why GP and GE can characterize quantum phase transitions, we need to consider the space formed by the Hamiltonian $H(\xi)$, whose unique ground state is given by $\ket{\psi_0(\xi)}$. Thus, the geometric tensor $\mathcal{T}_{\mu\nu}$ can be written as \cite{zanardi_informationtheoretic_2007, camposvenuti_quantum_2007}
\begin{eqnarray}\label{geo_tensor}
  \mathcal{T}_{\mu\nu} = \sum_{n\neq 0}\frac{\bra{\psi_0}\p_\mu H \ket{\psi_n}\bra{\psi_n}\p_\nu H\ket{\psi_0}}{(E_0 - E_n)^2},
\end{eqnarray}
where $\ket{\psi_n}$ is the $n$th eigenstate of $H(\xi)$ with eigenvalue $E_n(\xi)$. Quantum phase transitions occurs at level crossing or avoided level crossing, and the geometric tensor $\mathcal{T}_{\mu\nu}$ in the form of Eq.(\ref{geo_tensor}) can capture this level crossing, as it display a singular behavior when level crossing happens. Since GP and GE are related to the geometric tensor through Eqs.(\ref{ge_tensor}) and (\ref{gp_tensor}), they can also effectively capture quantum phase transitions.

\section{GP and GE in an interferometry setup}\label{sec:interf}
In addition to their unified description from the quantum geometry perspective, the GE and GP can also be conceptualized through an interferometry setup \cite{azimimousolou_unifying_2013}. This approach not only provides a deeper understanding but also offers a practical protocol for their measurement. The setup requires an auxiliary qubit, which in our case can be another Rydberg atom with states $\ket{g_0}, \ket{r_0}$. This auxiliary qubit serves as the two arms of the interferometer. Utilizing this ancilla, we prepare the initial state as:
\begin{eqnarray}
|\Psi_i\rangle = \frac{1}{\sqrt{2}}(|r_0\rangle + |g_0\rangle) \otimes |\psi\rangle,
\end{eqnarray}
where $\ket{\psi}$ is the ground state of the Rydberg atom chain under consideration.

This interferometer protocol consists of the following steps:

1. In the $\ket{r_0}$ arm, project the state $\ket{\psi}$ onto its closest product state $\ket{\psi_p}$. This can be achieved by applying a series of two-qubit operations: 
\begin{eqnarray}
  O_1 = \prod_i \left[\ket{g_0}\bra{g_0}\otimes I^i + \ket{r_0}\bra{r_0}\otimes \ket{\psi_p^i}\bra{\psi_p^i}\right],
\end{eqnarray}
where $\ket{\psi_p}$ is a product state that can be written as $\ket{\psi_p} = \otimes_i \ket{\psi_p^i}$, and $I^i$ is the identify operator on site $i$ of the Rydberg chain. The operation $\ket{g_0}\bra{g_0}\otimes I^i + \ket{r_0}\bra{r_0}\otimes \ket{\psi_p^i}\bra{\psi_p^i}$ is a two-qubit operation on the ancilla qubit and site $i$ of the Rydberg chain.

2. Apply the unitary operators:
\begin{eqnarray}
U_\psi = \otimes_i e^{\frac{i}{2}\pi[\hat{\sigma}_i^z - \bra{\psi}\hat{\sigma}_i^z\ket{\psi}]}
\end{eqnarray}
and
\begin{eqnarray}
U_{\psi_p} = \otimes_i e^{\frac{i}{2}\pi\hat{\sigma}_i^z}
\end{eqnarray}
to the state $\ket{\psi}$ and the product state $\ket{\psi_p}$ respectively. This is equivalent to applying the following set of two-qubit operations to the state $O_1\ket{\Psi_i}$
\begin{eqnarray}
  O_2 = \prod_i\left[\ket{g_0}\bra{g_0}\otimes e^{\frac{i}{2}\pi[\hat{\sigma}_i^z - \bra{\psi}\hat{\sigma}_i^z\ket{\psi}]} + \ket{r_0}\bra{r_0}\otimes e^{\frac{i}{2}\pi\hat{\sigma}_i^z} \right].
\end{eqnarray}

3. Apply a $U(1)$ phase shift $e^{-if}$ to the $\ket{g_0}$ arm using the operator: 

\begin{eqnarray}
  O_3 = \prod_i \left[\ket{g_0}\bra{g_0}\otimes e^{-if}I^i + \ket{r_0}\bra{r_0}\otimes I^i\right].
\end{eqnarray}

4. After applying the previous three steps, the state becomes:
\begin{eqnarray}
  |\Psi_f\rangle = \frac{1}{\sqrt{2}}|g_0\rangle\otimes U_\psi|\psi\rangle + \frac{1}{\sqrt{2}}|r_0\rangle \otimes U_{\psi_p}|\psi_p\rangle\langle\psi_p|\psi\rangle.
\end{eqnarray}
By defining $|\pm\rangle = \frac{1}{\sqrt{2}}(|g_0\rangle\pm |r_0\rangle)$, this state can be rewritten as 
\begin{eqnarray}
  |\Psi_f\rangle = \frac{1}{2} |+\rangle\otimes \left(|\Psi_1\rangle + |\Psi_2\rangle \right) + \frac{1}{2}|-\rangle\otimes \left(|\Psi_1\rangle - |\Psi_2\rangle\right)
\end{eqnarray}
where $|\Psi_1\rangle = U_\psi|\psi\rangle, |\Psi_2\rangle = U_{\psi_p}|\psi_p\rangle\langle\psi_p|\psi\rangle$. 
By post-selecting on the $|+\rangle$ state of the ancilla qubit, we obtain the desired state $(|\Psi_1\rangle + |\Psi_2\rangle)/2$. 

The resulting interference fringe is given by:
\begin{eqnarray}
I = \frac{1}{4}\left[1 + |\inner{\psi_p}{\psi}|^2\right] + \frac{1}{2}\mathrm{Re}(A e^{-if})
\end{eqnarray}
where
\begin{eqnarray}
A &=& \bra{\psi}U_{\psi}^\dag U_{\psi_p}\ket{\psi_p}\inner{\psi_p}{\psi} \nn\\
&=& |\inner{\psi_p}{\psi}|^2 e^{i\Phi_B}
\end{eqnarray}
Based on these results, we can define a complex-valued GE as:
\begin{eqnarray}
E_c(\psi) = -\log_2 A = E(\psi) - i\frac{\Phi_B}{\ln 2},
\end{eqnarray}
where the real part represents the geometric entanglement $E(\psi)$ and the imaginary part corresponds to the geometric phase. This formulation elegantly combines both the GE and GP into a single complex-valued quantity, providing a unified measure of these quantum properties.

It is worth noting that our protocol primarily relies on two-qubit operations. Given that high-fidelity two-qubit gates have already been successfully implemented in Rydberg atom arrays \cite{ebadi_quantum_2021,evered_highfidelity_2023,levine_parallel_2019}, we believe this protocol is within the reach of current experimental techniques.

\section{Conclusions}\label{sec:conclusion}
In this study, we have demonstrated the effectiveness of GP and GE as sensitive probes of quantum criticality in Rydberg atom chains. Through DMRG calculations and finite-size scaling analysis, we have characterized the critical properties of transitions from disordered to $Z_2$ and $Z_3$ ordered phases.

Our results show that GP and GE exhibit distinct signatures at quantum phase transitions, providing complementary information about the system's critical properties. We have explored a unifying description of GP and GE from a quantum geometric perspective, deepening our understanding of the connection between these quantities and the critical behavior of Rydberg atom chains. We have also discussed the potential adaptation of an interferometric setup for the experimental measurement of GP and GE in these systems. While challenges remain, this approach could enable future experimental verification of our theoretical predictions.

A promising direction for future research is the investigation of the Uhlmann phase \cite{uhlmann_parallel_1986, uhlmann_gauge_1991, uhlmann_density_1993, ericsson_mixed_2003, ericsson_generalization_2003, wang_uhlmann_2024}, which extends the concept of geometric phase to finite temperatures and open quantum systems. This could provide insights into the behavior of Rydberg atom systems under more realistic conditions. Meanwhile, geometric entanglement can be generalized to mixed states in two ways: (i) by replacing the overlap between pure states in Eq. (\ref{lambda_max}) with the fidelity between two density matrices, and (ii) by purifying the mixed state to a pure state and calculating the GE for this pure state. Both approaches introduce additional complexity to the calculation of GE and need further investigation.

In conclusion, our study highlights the power of geometric concepts in characterizing quantum phase transitions in Rydberg atom chains. As quantum simulation platforms advance, we anticipate that these approaches will play an increasingly important role in our understanding of quantum many-body systems, opening new avenues for exploring fundamental questions in quantum physics and advancing quantum technologies.

\begin{acknowledgements}
CYW is supported by the Shuimu Tsinghua scholar program at Tsinghua University.
\end{acknowledgements}

\appendix

\section{Finite-size scaling analysis} \label{fss}
In this appendix, we provide relevant information about the finite-size scaling analysis used in our paper. Further details can be found in \cite{barber_phase_1983}.

For an infinite Rydberg atomic chain, the correlation length $ \xi $ diverges near the critical point $ \tilde{\delta}_c $ as:
\begin{eqnarray}
\xi \sim |\tilde{\delta} - \tilde{\delta}_c|^{-\nu},
\end{eqnarray}
where $ \nu $ is the critical exponent. If the derivative of the geometric phase $ \phi_B' $ scales logarithmically with the correlation length in the thermodynamic limit, we can write:
\begin{eqnarray}
\phi_B' = \kappa_{1,\phi'} \ln\xi + \text{const.}.
\end{eqnarray}
Substituting the divergence of $ \xi $, we obtain:
\begin{eqnarray}
\phi_B' = -\kappa_{2,\phi'} \ln |\tilde{\delta} - \tilde{\delta}_c| + \text{const.},
\end{eqnarray}
where $ \kappa_{2,\phi'} = -\nu \kappa_{1,\phi'} $. From this, it follows that:
\begin{eqnarray}
\nu = \left| \frac{\kappa_{2,\phi'}}{\kappa_{1,\phi'}} \right|.
\end{eqnarray}

In a finite-size system, if the system size $ L \ll \xi $, the correlation length is effectively limited by the system size. As a result, if $ \phi_B' $ scales logarithmically with $ \xi $, we expect it to scale with the system size $ L $ as:
\begin{eqnarray}
\phi_B' = \kappa_{1,\phi'} \ln L + \text{const.}.
\end{eqnarray}
Conversely, observing a logarithmic dependence of $ \phi_B' $ on the system size $ L $ in a finite system implies a logarithmic dependence on the correlation length $ \xi $ and, consequently, on $ |\tilde{\delta} - \tilde{\delta}_c| $.

\section{Quality function used for the finite size scaling}\label{app:quality}
In this appendix, we will briefly review the quality function that measures how well data collapse in the finite size scaling. This quality function is first introduced in Ref. \cite{houdayer_lowtemperature_2004}, which refined the one introduced in Ref. \cite{kawashima_critical_1993}. In the following, we use the GP density derivative $\phi_B'$ as an example to define the quality function. One can also follow the same procedure to define the quality function for the GE density derivative. 

We let $i$ labels the different system sizes $L_i$, $i = 1,2,..., k$, and $j$ labels the tuning parameters $\td_j$, $j = 1,2,...,n$, with $\td_1 < \td_2 < \cdots < \td_n$. Then, we define the scaled data as 
\begin{eqnarray}
  x_{ij} &=& L_i^{1/\nu}(\td_j - \td_c), \\
  y_{ij} &=& 1 - \exp[\phi_B'(i,j) - \phi_{B,m}'(i)].
\end{eqnarray}
And $dy_{ij}$ is the standard error of $y_{ij}$. Then, with this set of data $\{x_{ij}, y_{ij}, dy_{ij}\}$, we can define $Y_{ij}$ and $dY_{ij}$ which are the estimated position and the standard error of the master curve, which is the curve that all the data should collapse to, at $x_{ij}$. 

$Y_{ij}$ and $dY_{ij}$ at $x_{ij}$ are defined as follows: For each $i'\neq i$, we select two points $(x_{i'j'}, y_{i'j'}, dy_{i'j'})$ and $(x_{i',j'+1}, y_{i',j'+1}, dy_{i',j'+1})$ that satisfy $x_{i'j'} \leq x_{ij} \leq x_{i',j'+1}$. If there is no such points for this $i'$ data set, we do not select any points from this set. After doing such selections, suppose we have $m$ data points $(x_l, y_l, dy_l)$, $l = 1,2,...,m$. With this data points, we can do a weighted least squares linear fit. Then, $Y_{ij}$ is the value of that fitted straight line at $x_{ij}$, and $dY_{ij}$ is the associated standard error, i.e.
\begin{eqnarray}
  Y_{ij} &=& \frac{K_{xx}K_y-K_x K_{xy}}{\Delta} + x_{ij} \frac{KK_{xy}-K_x K_y}{\Delta}, \\
  dY_{ij} &=& \frac{1}{\Delta}(K_{xx} - 2x_{ij} K_x + x_{ij}^2 K)
\end{eqnarray}
where $w_l = \sum 1/dy_l^2,\ K = \sum w_l,\ K_x = \sum w_l x_l,\ K_y = \sum w_l y_l,\ K_{xx} = \sum w_l x_l^2,\ K_{xy} = \sum w_l x_l y_l$, and $\Delta = K K_{xx} - K_x^2$.
Then we can define the quality function is defined as the mean square distance of the data set to the fitted master curve in units of standard errors, i.e.
\begin{eqnarray}
  S(\td_c,\nu) = \frac{1}{N}\sum_{ij}\frac{(y_{ij} - Y_{ij})^2}{dy_{ij}^2 + dY_{ij}^2}.
\end{eqnarray}
By minimizing this quality function over all possible $\td_c$ and $\nu$, we should get the correct critical point and critical exponent.

\bibliography{ref}

\end{document}